\documentclass[aps,prl,reprint,amsmath,amssymb,longbibliography,nofootinbib,floatfix,superscriptaddress]{revtex4-1}
\usepackage{graphicx}
\usepackage{physics}
\usepackage{xcolor}
\usepackage{comment}
\usepackage{makecell}
\usepackage{hyperref}
\usepackage{bm}
\usepackage{letltxmacro}

\AtBeginDocument{%
    \newwrite\bibnotes
    \def\bibnotesext{Notes.bib}
    \immediate\openout\bibnotes=\jobname\bibnotesext
    \immediate\write\bibnotes{@CONTROL{REVTEX41Control}}
    \immediate\write\bibnotes{@CONTROL{%
    apsrev41Control,author="08",editor="1",pages="1",title="0",year="1", doi="0"}}
     \if@filesw
     \immediate\write\@auxout{\string\citation{apsrev41Control}}%
    \fi
}%

\DeclareMathOperator{\sinc}{sinc}

\renewcommand{\Im}{\imaginary}

\providecommand{\ee}{\text{e}}

\usepackage[dvipsnames]{xcolor}

\newcommand{\unit}[1]{\mathbf{\hat{#1}}}
\LetLtxMacro{\originaleqref}{\eqref}
\renewcommand{\eqref}{Eq.~\originaleqref}

\date[]{}
\begin{document}

\title{Time diffraction of optical helicity}

\author{Alex J. Vernon}
\email{alexvernon10@gmail.com}
\affiliation{Department of Physics and London Centre for Nanotechnology, King's College London, Strand, London WC2R 2LS, UK}
\affiliation{Donostia International Physics Center (DIPC), Donostia-San Sebasti\'an 20018, Spain}

\author{Jingyi Wu}
\affiliation{Department of Physics and London Centre for Nanotechnology, King's College London, Strand, London WC2R 2LS, UK}

\author{Anton Y. Bykov}
\affiliation{Nanophotonics Centre, Cavendish Laboratory, Department of Physics, University of Cambridge, CB3 0US, UK}
\affiliation{Department of Physics and London Centre for Nanotechnology, King's College London, Strand, London WC2R 2LS, UK}

\author{Guy L. Whitworth}
\affiliation{Department of Physics and London Centre for Nanotechnology, King's College London, Strand, London WC2R 2LS, UK}

\author{Henry Cossey}
\affiliation{Department of Physics and London Centre for Nanotechnology, King's College London, Strand, London WC2R 2LS, UK}

\author{\\Francisco~J. Rodr\'iguez-Fortu\~no}
\email{francisco.rodriguez\_fortuno@kcl.ac.uk}
\affiliation{Department of Physics and London Centre for Nanotechnology, King's College London, Strand, London WC2R 2LS, UK}

\author{Anatoly V. Zayats}
\email{a.zayats@kcl.ac.uk}
\affiliation{Department of Physics and London Centre for Nanotechnology, King's College London, Strand, London WC2R 2LS, UK}

\begin{abstract}
If in a classic double-slit interference experiment the light that passes though one slit is in an orthogonal polarisation state to that of the second slit, no visible interference fringes emerge.
Interference fringes are instead present in a different observable that relates to the difference in energy carried by two orthogonal polarisation components, causing spatially varying polarisation patterns: the optical helicity density in the case of linearly polarised fields.
Here, we develop a generalised treatment of interference in space and time, and experimentally demonstrate the helicity patterns produced by interference of orthogonally polarised pulses, which would be produced by two temporal slits in time diffraction observations.
The developed approach may be important for understanding and applications of complex polarisation patterns and polarisation effects in time varying media. 
\end{abstract}

\maketitle


{\it Introduction.---}
Wave interference is fundamental to both classical and quantum physics. Among its various manifestations, the double-slit experiment has, since the early nineteenth century, underpinned our understanding of the wave nature of light and coherence, prefiguring modern optical technologies such as precision metrology \cite{Asakura1981,Meynart1983,Lavery2013,Cheng2025} and structured light generation \cite{Bliokh2015,Rubinsztein-Dunlop2017,Forbes2021}, as well as many complex phenomena that can emerge in the interference of as few as two plane waves \cite{Tang2010,Bliokh2014,Bekshaev2015,vanKruining2018,Isaule2022,Kilianski2023}.
In quantum technologies, two-slit interference demonstrates quantum superposition, establishes the role of information (the knowledge of ‘which path’) and is a basis for quantum entanglement experiments \cite{Scully1991}.
Numerous variations of the experiment in electromagnetic and atom optics have been performed \cite{Pfleegor1967,Carnal1991,Walborn2002,Bach2013}, including demonstrations of a quantum eraser and a delayed quantum eraser in a double-slit configuration \cite{Scully1991}, and interference from slits separated in time instead of in space \cite{Tirole2023}.
New interpretations of the formation of interference fringes have been recently proposed based on the superposition of ‘bright’ and ‘dark’ quantum optical states without involving the wave nature of light \cite{VillasBoas}.  

In a textbook explanation, the light at each slit must be coherent and in the same polarisation state in order to produce an interference pattern: the interference of orthogonal linear or circularly polarised light beams will not produce fringes.
However, in this case, while there are no fringes in the intensity of the field, fringes in the difference in energy density carried by orthogonal polarisation components are still present, which can be revealed by placing an appropriate polariser after each slit.
This observation is responsible for spatially varying polarisation patterns \cite{Setl2006,Setl2006_2,Hannonen2020,Banguilan2023,Aita2025} and is applied in quantum eraser experiments \cite{Walborn2002}.

\begin{figure}[b!]
    \centering
    \includegraphics[width=\columnwidth]{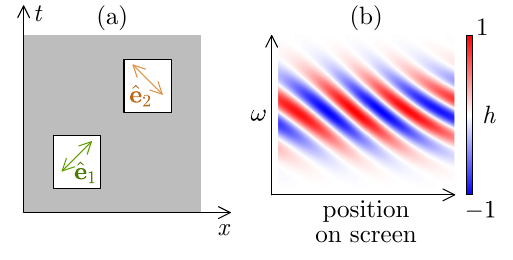}
    \caption{General principle of space and time diffraction of optical helicity.
    (a) Schematics of two slits in an opaque medium separated along the $x$ axis appearing at different times.
    Linearly polarised light incident on these slits has orthogonal polarisations after passing the slits.
    (b) Schematic of the resultant interference fringes in the far field helicity density, appearing both as a spatial pattern on a screen and in the field frequency spectrum $\omega$.}
    \label{fig1}
\end{figure}

If the two beams have orthogonal linear polarisations, interference fringes appear in the electromagnetic helicity density \cite{Lipkin1964,Calkin1965,Trueba1996,Afanasiev1996,Bliokh2011,Cameron2012,Cameron2012_2,Bliokh2013} given for monochromatic light by
\begin{equation}\label{helicity}
    h=-\frac{1}{2\omega c}\Im\{\mathbf{E}^*\cdot\mathbf{H}\}\,,
\end{equation}
where $\mathbf{E}$ and $\mathbf{H}$ are complex phasors for the electric and magnetic fields.
This helicity density relates to the difference in number density of left- and right-handed photons \cite{Calkin1965,Trueba1996,Afanasiev1996,Bliokh2011,Cameron2012}.
Spatially varying helicity patterns, or helicity lattices, can be synthesised in general by superimposing linear or circularly orthogonally polarised plane waves and have been proposed as a method for sorting chiral enantiomers \cite{vanKruining2018,Isaule2022,Kilianski2023} or generating topologically protected polarisation structures from interference of orthogonally polarised beams with different optical angular momentum \cite{advphot-yijie}.

With recent advances in temporal modulation of materials \cite{Engheta2023}, many fundamental electromagnetic phenomena, photonic systems, and cornerstone experiments like Young's slits \cite{Tirole2023}, have now been transformed from the spatial to the temporal domain \cite{Guo2019,Zhou2020,Pacheco-Pea2021,Galiffi2022,Moussa2023,Bahrami2025,Stefanini2025,Harwood2025,Raziman2025}. In temporal two-slit interference, the refractive index of a spatially uniform material is changed instantaneously by two laser pulses with a delay between them to create temporal interfaces that act as slits separated in time \cite{Tirole2023}, similarly to double spatially separated slits, producing interference fringes in the frequency spectrum of energy density.

In this work, we provide a generalised treatment of two beam interference in space and time for orthogonally polarised electromagnetic fields. We experimentally demonstrate interference of orthogonally polarised light beams in the helicity density observable both in spatial two-slit configuration and in its temporal realisation. The former is achieved with two slits acting as orthogonally oriented linear polarisers. The latter emulates the amplitude-modulated signal that would be created by an interaction of light with slits that appear in time: two orthogonal, linearly polarised pulses with a delay between them.

{\it Invisible interference fringes.---}The basic feature of any two-slit diffraction experiment---over space or in time---is a far field described by
\begin{equation}\label{E_interference}
    \mathbf{E}(\alpha)=\left(\unit{e}_1\ee^{-i\phi(\alpha)/2}+\unit{e}_2\ee^{i\phi(\alpha)/2}\right)A(\alpha)\,,
\end{equation}
where $\unit{e}_1$ and $\unit{e}_2$ are the polarisation states of the field at each slit.
A phase offset $\phi$ between these polarisation vectors varies according to a parameter $\alpha$.
In Young's original double-slit experiment, $\alpha$ corresponds to the horizontal position on the screen capturing the light from the two slits, while in temporal two-slit interference $\alpha$ corresponds to angular frequency.
An envelope function $A(\alpha)$ ($\sim\sinc{\alpha}$ for rectangular slits) characterises how amplitude interference fringes diminish with distance from a central maximum.
\begin{figure}[t!]
    \centering
    \includegraphics[width=\columnwidth]{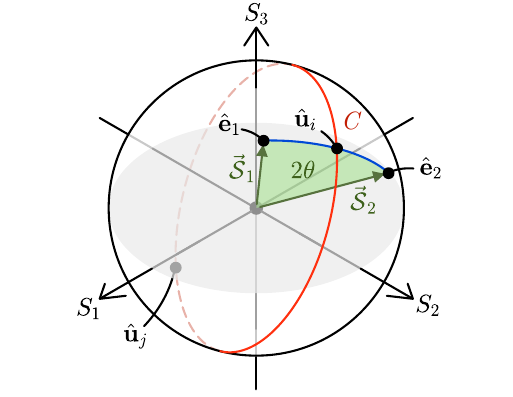}
    \caption{Geometry of arbitrary polarisation states of the two interfering fields $\unit{e}_1$ and $\unit{e}_2$ on the Poincar\'e sphere.
    The fields are characterised by the Stokes vectors (green arrows) $\vec{\mathcal{S}}_1$ and $\vec{\mathcal{S}}_2$, respectively, which subtend an angle $2\theta$ that is related to the inner product of the two unit polarisation vectors via $|\unit{e}_1^*\cdot \unit{e}_2|=\cos\theta$.
    The shortest geodesic connecting $\unit{e}_1$ and $\unit{e}_2$ is shown in blue.
    A great circle $C$ (red) bisects this geodesic such that all points on $C$ are equidistant from $\unit{e}_1$ and $\unit{e}_2$.
    Two antipodal points on this great circle, corresponding to $\unit{u}_i$ and $\unit{u}_j$, can be chosen.
    In two-beam interference, the scalar $|\mathbf{E}^*\cdot\unit{u}_i|^2-|\mathbf{E}^*\cdot\unit{u}_j|^2$ contains fringes.}
    \label{fig2}
\end{figure}

A general description of interference fringes that emerge in \eqref{E_interference}, in both space and time, can be approached by considering the geometry of the problem on the Poincar\'e sphere (Fig.~\ref{fig2}).
The polarisation state of light at each slit is assigned a Stokes vector, $\vec{\mathcal{S}}_1=\unit{e}_1^*\cdot\left(\vec{\hat{\sigma}}\right)\unit{e}_1$ and $\vec{\mathcal{S}}_2=\unit{e}_2^*\cdot\left(\vec{\hat{\sigma}}\right)\unit{e}_2$, where $\vec{\hat{\sigma}}$ is the vector of Pauli matrices, each pointing to different locations on the sphere surface.
Whatever their orientation, the tips of $\vec{\mathcal{S}}_1$ and $\vec{\mathcal{S}}_2$ can be connected by two geodesics, the shorter of the two (Fig.~\ref{fig2}) having a length of $2\theta$.
Bisecting both of these geodesics is a great circle $C$, on which every point is equidistant between the tips of $\vec{\mathcal{S}}_1$ and $\vec{\mathcal{S}}_2$ (Fig.~\ref{fig2}).
The inner product of the polarisation states relates to the Stokes vectors via $|\unit{e}_1^*\cdot\unit{e}_2|^2=(1+\vec{\mathcal{S}}_1\cdot\vec{\mathcal{S}}_2)/2$, meaning $|\unit{e}_1^*\cdot\unit{e}_2|=\cos\theta$.
As such $\theta=0$ and $\pi$ corresponds to co-polarised fields, while $\theta=\pi/2$ corresponds to orthogonally polarised ones.
Any two antipodal points on $C$ can be taken to correspond to basis states $\unit{u}_i$ and $\unit{u}_j$ (where $\unit{u}_i^*\cdot\unit{u}_j=0$).
We choose these antipodal points on $C$ to lie in the same plane as $\vec{\mathcal{S}}_1$ and $\vec{\mathcal{S}}_2$.

The square magnitude (intensity) of the interference field in \eqref{E_interference} can be calculated as a function of $\theta$ and $\phi$, giving
\begin{equation}\label{Esq}
    |\mathbf{E}|^2=(1+\cos\theta\cos\phi)2A^2\,,
\end{equation}
which is true up to a global phase offset $\beta$ between the fields that influences $\phi\to\phi+\beta$.
If the polarisation states of both fields are the same and in-phase, $\theta=0$, then $|\mathbf{E}|^2 = (1+\cos \phi)2A^2$, producing interference fringes governed by the phase offset $\phi(\alpha)$.
Should the polarisation states at the slits be orthogonal, $\theta=\pi/2$ ($\unit{e}_1^*\cdot\unit{e}_2=0$), then $|\mathbf{E}|^2 = 2A^2$ for all $\phi(\alpha)$ and the double-slit diffraction pattern vanishes.

Yet interference fringes resurface in other observables.
The field intensity in \eqref{Esq} can be separated into contributions from the electric field orthogonal $\unit{u}_i$ and $\unit{u}_j$ components, that is $|\mathbf{E}|^2=|\mathbf{E}^*\cdot\unit{u}_i|^2+|\mathbf{E}^*\cdot\unit{u}_j|^2=|E_i|^2+|E_j|^2$.
Unlike their sum in \eqref{Esq}, the {\it difference} between the two contributions,
\begin{equation}\label{difference}
    |E_i|^2-|E_j|^2=(\cos\theta+\cos\phi)2A^2\,,
\end{equation}
contains interference fringes over $\phi$ {\it independently} of the polarisation states of interfering fields.
Therefore, a diffraction pattern is always measurable in any coherent two-beam interference experiment, spatial or temporal, and irrespective of the fields polarisation, so long as suitable analysers are placed after the slits to detect the correct quantity.
Equations (\ref{Esq}) and (\ref{difference})  are correct up to the global phase offset $\beta$ that translates the fringes along the $\alpha$ axis.

The physical significance of the quantity represented by \eqref{difference} depends on $\unit{u}_i$ and $\unit{u}_j$.
In this work, we are concerned with optical helicity density, which in paraxial light is proportional to $|E_+|^2-|E_-|^2$, i.e. the difference in the intensities of right- and left-handed polarisation components, which have the basis vectors $\unit{u}_\pm=(\unit{x}\pm i\unit{y})/\sqrt{2}$.
This relation follows from the fact that the magnetic field of a paraxial wave, propagating along $\unit{z}$, is $\mathbf{H}=\sqrt{\varepsilon_0/\mu_0}\,\unit{z}\times\mathbf{E}$, while $\unit{z}\times\unit{u}_\pm=\mp i\unit{u}_\pm$.
Evaluating \eqref{helicity} with $\mathbf{E}=E_+\unit{u}_++E_-\unit{u}_-$ gives $h=(|E_+|^2-|E_-|^2)\varepsilon_0/(2\omega)$.

A helicity diffraction pattern with alternating sign and a fringe-free field intensity can be achieved with suitable orthogonally polarised fields.
Taking $\unit{u}_i=\unit{u}_+$ and $\unit{u}_j=\unit{u}_-$ to correspond to the north and south poles of the Poincar\'e sphere (Fig.~\ref{fig2}), orthogonality $\theta=\pi/2$ implies that $\unit{e}_1$ and $\unit{e}_2$ must lie on the sphere equator.
According to \eqref{difference}, we should expect alternating fringes of positive and negative helicity in $\phi$ space.
Choosing, for instance, $\unit{e}_1=\unit{x}$ and $\unit{e}_2=\unit{y}$ to be horizontally and vertically polarised, respectively, gives $h=\varepsilon_0\omega^{-1}A^2\sin\phi$, while for diagonally and anti-diagonally polarised fields, $\unit{e}_1=(\unit{x}+\unit{y})/\sqrt{2}$ and $\unit{e}_2=(\unit{x}-\unit{y})/\sqrt{2}$, $h=-\varepsilon_0\omega^{-1}A^2\sin\phi$ is obtained.
Equations (\ref{E_interference})--(\ref{difference}) can be directly applied to both spatial and temporal two-slit interference experiments with orthogonally polarised fields.

{\it Space diffraction.---}Before measuring the diffraction of helicity in time, we experimentally demonstrate the principle in classic spatial two-slit interference.

\begin{figure}[t!]
    \centering
    \includegraphics[width=\columnwidth]{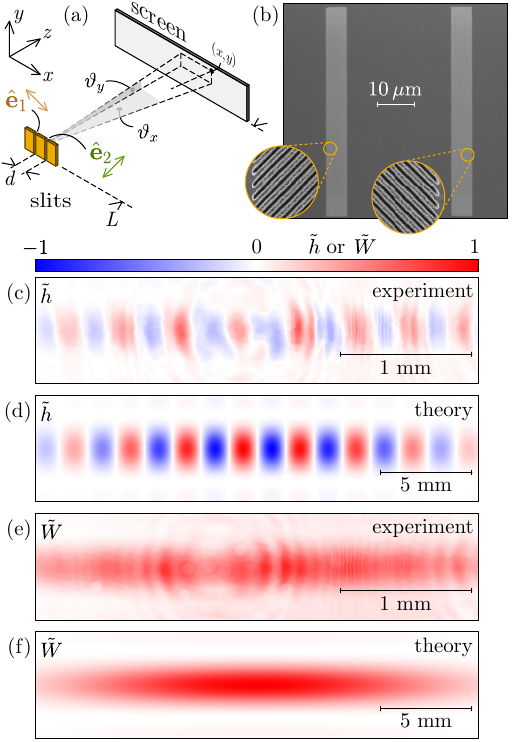}
    \caption{Experimental observation of helicity diffraction pattern with the slits separated in space.
    (a) A schematic of the two-slit interference experiment with labelled parameters.
    (b) SEM images of the fabricated slits, each $5\,\mu{\rm m}$ wide, which are formed of periodic diagonal slots that polarise transmitted light diagonally for one slit and antidiagonally for another.
    (c) Experimentally measured helicity interference pattern.
    (d) Simulated interference pattern using the electric field of \eqref{E_interference} and experimental parameters for a distance $L=0.1\,$m to a screen.
    Note that the $5\,$mm scale bar corresponds to the free-space pattern at $L=0.1\,$m, while the $1\,$mm scale bar in (c) corresponds to the pattern after focusing onto the camera sensor.
    (e) Experimental and (f) simulated distributions of normalised energy density.}
    \label{fig3}
\end{figure}

In a simple two-slit interference experiment (Fig.~\ref{fig3}(a)), the slits lie in the $xy$ plane, are of finite width $a$ and height $b$ and are separated along the $x$ axis by a distance $d$.
They are placed before a screen at a distance $L$ along the $z$ axis.
The electric field phasor on the screen has the form \eqref{E_interference} with $\alpha\to x$.
The horizontal $x$ and vertical $y$ position on the screen define the angles $\vartheta_x=\arctan x/L$ and $\vartheta_y=\arctan y/L$ (Fig.~\ref{fig3}(a)) which together determine the phase offset $\phi=kd\sin\vartheta_x$ and envelope function $A=\sinc(\pi a\lambda^{-1}\sin\vartheta_x)\sinc(\pi b\lambda^{-1}\sin\vartheta_y)$. 
We experimentally measured a helicity diffraction pattern with a vertically polarised laser beam of $800$ nm wavelength incident on two polarising slits (Fig.~\ref{fig3}(b)). The slits (a width of $a=5\,\mu\text{m}$, a length of $b=50\,\mu\text{m}$ and a separation of $d=30\,\mu\text{m}$) were fabricated in a 115-nm-thick gold film, sputtered on glass with a 10~nm Ta$_2$O$_5$ adhesion layer. The grating pattern was defined using electron beam lithography, followed by wide-beam ion-milling at normal incidence in order to etch the grating through the gold film. The slits were designed to polarise transmitted light diagonally (45$^\circ$ to the slit length) and anti-diagonally (-45$^\circ$) using a grating of period of 200 nm with a slot filling fraction of 40\%, with appropriate orientation with respect to the slit length (see the insets of Fig.~\ref{fig3}(b)).
While no significant fringes are visible in the intensity $\tilde{W}=S_0/S_{0\text{max}}$ measurements (Fig.~\ref{fig3}(e,f)), the diffraction pattern is obvious in the helicity distribution, which was measured by placing a wave plate and a polariser for evaluating the Stokes parameter $S_3$, which relates to the normalised helicity as $\tilde{h}\equiv S_3/S_{0\text{max}}$, where $S_{0\text{max}}$ is the maximum measured value of $S_0$.

\begin{figure*}[t]
    \centering
    \includegraphics[width=\textwidth]{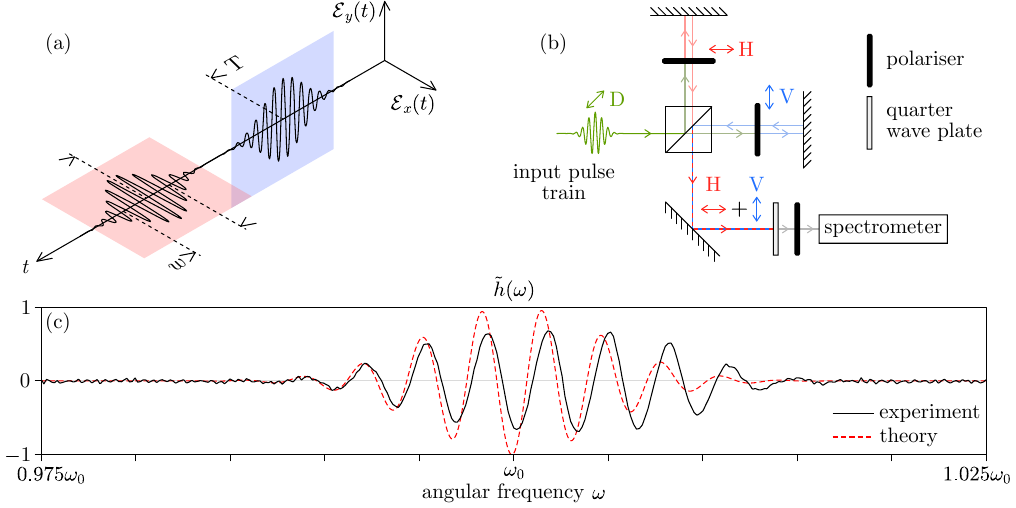}
    \caption{Experimental measurement of helicity time diffraction.
    (a) Illustration of two orthogonally polarised pulses of light, each with amplitude FWHM of $w=150$ fs and a central wavelength of 800~nm, separated in time by a delay $T$.
    (b) Experimental setup, consisting of a train of diagonally polarised pulses entering a Michelson interferometer, with a horizontal (H) or vertical (V) polariser in each arm.
    The recombined light leaving the beam splitter is a train of H/V pulses.
    A quarter wave plate and a output polariser allow the intensity of RHCP and LHCP components of the recombined pulse train to be measured as a function of wavelength.
    (c) Measured (black line) and simulated (red dashed line) diffraction patterns in angular frequency space of the normalised helicity for pulses with the experimental parameters. The pattern is normalised by the peak energy density $W$ at $\omega=\omega_0$, i.e., $\tilde{h}(\omega)=\omega h(\omega)/W(\omega_0)\equiv S_3(\omega)/S_0(\omega_0)$, where $S_3$ and $S_0$ are the Stokes parameters.}
    \label{fig4}
\end{figure*}

{\it Time diffraction.---}Whereas spatial diffraction involves spatially localised sources of light interfering in real space, time-diffraction concerns temporally localised light pulses that interfere in frequency space.
We consider two pulses with identical spectra, polarised along unit vectors $\unit{e}_1$ and $\unit{e}_2$ propagating in the same direction with a time separation of $T$ (Fig.~\ref{fig4}(a)).
As a pulse can be described as a set of superimposed plane waves, each of frequency $\omega$ and with an amplitude spectrum $A(\omega)$ determined by the pulse duration, the combined electric field of the overlapping pulses is identical to \eqref{E_interference} with $\alpha$ replaced by $\omega$.
The frequency-dependent phase offset $\phi(\omega)=-\omega T$ is created by the temporal separation of the pulses.
The instantaneous electric field vector is the Fourier transform of \eqref{E_interference}, i.e., $\boldsymbol{\mathcal{E}}(t)=\int_{-\infty}^{\infty}\mathbf{E}(\omega)\exp(-i\omega t)d\omega$ (Fig.~\ref{fig4}(a)).

Should $\unit{e}_1=\unit{x}$ and $\unit{e}_2=\unit{y}$, frequencies for which $\omega T=n\pi$ ($n\in\mathbb{Z}$) construct a diagonally polarised field that carries zero helicity density, while frequencies satisfying $\omega T=2n\pi\pm\pi/2$ construct right/left handed circularly polarised field with maximal helicity density.
Helicity density is smoothly distributed among intermediate frequencies in a characteristic fringe pattern over the $\omega$ axis that alternates in sign according to $h=\varepsilon_0\omega^{-1}A^2(\omega)\sin{\omega T}$.
It is important to highlight that the pulses {\it do not overlap in time} and, being only $\unit{x}$ or $\unit{y}$ polarised, {\it do not individually carry helicity}.

In order to experimentally measure a temporal helicity diffraction pattern, a train of diagonally polarised pulses was used as an input into a Michelson interferometer. Additional polarisers were placed in each interferometer arm to produce orthogonal states of polarisation (Fig.~\ref{fig4}(b)).
The pulses have a Gaussian shape in time with an amplitude full width at half maximum (FWHM) of $w=150$ fs and a central frequency of $\omega_0=2.355$ PHz ($\lambda_0=800$ nm); these parameters correspond to the Gaussian envelope $A(\omega)=A_0\exp(-(\omega-\omega_0)^2w^2/(16\ln2))$, where $A_0=w/(4\sqrt{\pi\ln2})$ for a pulse with unit amplitude.
Linear polarisers were introduced in each arm of the interferometer: oriented horizontally (H) in one arm and vertically (V) in the other, and the path difference between the two arms was adjusted so that in the recombined pulse train H and V pulses alternate, being separated by a delay $T$.
With a quarter wave plate (QWP) and a linear polariser at the output, the Stokes parameter $S_3$, normalised by peak intensity at 800~nm, was measured as a function of wavelength using a spectrometer (Fig.~\ref{fig4}(c)). The QWP converts the H/V pulse train into a right- and left-handed circularly polarised (RHCP and LHCP, respectively) pulse train, before it goes through the linear polariser, whose outputs are subtracted for $0^\circ$ and $90^\circ$ orientations.
The measurement of time-diffracted helicity interference $\tilde{h} = S_3(\omega)/S_0(\omega_0)$ in orthogonal linearly polarised pulses separated in time is equivalent to a measurement of the time diffraction interference of the normalised first Stokes parameter $S_1(\omega)/S_0(\omega_0)$ for orthogonal circularly polarised pulses separated in time.
The measured helicity pattern corresponds well to the simulated one given by $\tilde{h}=[A^2(\omega)/A^2(\omega_0)]\sin\omega T$ (Fig.~\ref{fig4}(c)).
Note that the fringes in helicity are extremely sensitive to the pulse delay $T$, which was determined by the theoretical fit to be 821.1 fs.
Therefore, temporal diffraction patterns could be used for precision distance and time measurements.

{\it Conclusion.---}We have shown theoretically and experimentally that, just as in two-slit spatial interference, hidden time-diffraction patterns can be uncovered even when the pulses of light that would be produced by time slits are orthogonally polarised.
Orthogonal, linearly polarised pulses, which do not individually carry helicity, create fringes of helicity density in frequency space due to interference.
This effect could form the basis for further temporal optics experiments such as a quantum eraser experiment in time.
These invisible fringes have physical implications in light-matter interactions as a spatial gradient in helicity density is one of several ways to bring about chiral optical forces \cite{Cameron2014,Hayat2015} and spatial helicity diffraction patterns (or helicity lattices) have been proposed as a way to separate chiral enantiomers \cite{vanKruining2018}.
A train of orthogonal, linearly polarised pulses, on the other hand, has a frequency-dependent chirality, so that the chiral forces exerted by the pulse train on an illuminated chiral particle would depend on the particle resonance wavelength: whether it lies in a positive or negative helicity fringe.
This could enable the precision tuning of chiral optical forces via the temporal pulse separation $T$.
Changing the orthogonal polarisation of the pulses, meanwhile, changes the quantity in \eqref{difference} that contains the interference fringes (for instance, right- and left-circular pulses would produce a time-diffraction pattern in $|E_x|^2-|E_y|^2$) which could couple to a different property of a particle (e.g., the orientation of a nanorod, which couples to polarisation components along its long axis).
That these fringes are extremely sensitive to the pulse separation time could provide a way to delicately tune optical forces in pulsed optical traps or other scenarios.

{\it Acknowledgments.} This work was supported by the Engineering and Physical Sciences Research Council (EPSRC) grant META4D (EP/Y015673). G.L.W. acknowledges support from the Horizon Europe Guarantee project EP/Z000912/1.

\bibliography{bibliography}

\end{document}